\documentclass[%
 reprint,
superscriptaddress,
 amsmath,amssymb,
 aps,
pra, 
longbibliography,
floatfix ]{revtex4-1}

\usepackage{natbib} 
\usepackage{amsmath} 
\usepackage{graphicx} 
\usepackage[usenames,dvipsnames]{color}  
\usepackage{longtable}
\usepackage{multirow}
\usepackage{amsmath}
\usepackage{epstopdf}

\usepackage[english]{babel}

\makeatletter
\def\bbl@set@language#1{%
  \edef\languagename{%
    \ifnum\escapechar=\expandafter`\string#1\@empty
    \else\string#1\@empty\fi}%
  \@ifundefined{babel@language@alias@\languagename}{}{%
    \edef\languagename{\@nameuse{babel@language@alias@\languagename}}%
  }%
  \select@language{\languagename}%
  \expandafter\ifx\csname date\languagename\endcsname\relax\else
    \if@filesw
      \protected@write\@auxout{}{\string\select@language{\languagename}}%
      \bbl@for\bbl@tempa\BabelContentsFiles{%
        \addtocontents{\bbl@tempa}{\xstring\select@language{\languagename}}}%
      \bbl@usehooks{write}{}%
    \fi
  \fi}
\newcommand{\DeclareLanguageAlias}[2]{%
  \global\@namedef{babel@language@alias@#1}{#2}%
} \makeatother

\DeclareLanguageAlias{en}{english}

\begin{document}

\title{Gender Fairness within the Force Concept Inventory}

\author{Adrienne Traxler}
\affiliation{Department of Physics, Wright State University,
Dayton, OH, 45345}
\author{Rachel Henderson}%
\author{John Stewart}
\author{Gay Stewart}
\affiliation{%
Department of Physics and Astronomy, West Virginia University,
Morgantown WV, 26506
}%
\author{Alexis Papak}
\affiliation{Department of Physics, University of Maryland,
College Park, MD, 20742}
\author{Rebecca Lindell}
\affiliation{Tiliadal STEM Education, West Lafayette, IN, 47907}
\date{\today}

\begin{abstract}

Research on the test structure of the Force Concept Inventory
(FCI) has largely  ignored gender, and research on FCI gender
effects (often reported as ``gender gaps'') has seldom
interrogated the structure of the test. These rarely-crossed
streams of research leave open the possibility that the FCI may
not be structurally valid across genders, particularly since many
reported results come from calculus-based courses where 75\% or
more of the students are men. We examine the FCI considering both
psychometrics and gender disaggregation (while acknowledging this
as a binary simplification), and find several problematic
questions whose removal decreases the apparent gender gap. We
analyze three samples (total $N_{pre}=5,391$, $N_{post}=5,769$)
looking for gender asymmetries using Classical Test Theory, Item
Response Theory, and Differential Item Functioning. The
combination of these methods highlights six items that appear
substantially unfair to women and two items biased in favor of
women. No single physical concept or prior experience unifies
these questions, but they are broadly consistent with problematic
items identified in previous research. Removing all significantly
gender-unfair items halves the gender gap in the main sample in
this study. We recommend that instructors using the FCI report the
reduced-instrument score as well as the 30-item score, and that
credit or other benefits to students not be assigned using the
biased items.

\end{abstract}

\maketitle

\section{Introduction}

The Force Concept Inventory (FCI)
has been studied using tools such as factor analysis
\citep{huffman_what_1995,scott2012exploratory}, item response theory
\citep{wang_analyzing_2010,morris2012item}, and network analysis
\citep{brewe_using_2016}. Though these investigations have probed
the structure and validity of the test, they have primarily
treated student data as a single undifferentiated sample and have
not studied gender effects. A largely separate branch of research
has explored gender differences in scores on the FCI and other
conceptual inventories. These studies have documented a ubiquitous
advantage for men on pretest questions, which often persists to
the posttest. Proposed explanations range from differences in
preparation, to instructional method (when examining gains), to
sociocultural factors such as stereotype threat. With some
exceptions, the literature on test construction largely ignores
gender effects, and the literature on gender effects focuses on
total score and takes the integrity of the instrument as a given.
Because a great deal of FCI data is collected from calculus-based
courses where 75\% or more of the students are male, it
remains an open question whether gender-blind validations of the
FCI for ``all students'' are in fact applicable to all, or whether
poorly-functioning items for women might be hidden in the
unbalanced sample.

In this paper, gender fairness will be explored in three samples
of FCI pretest and posttest data (total $N_{pre}=5,391$,
$N_{post}=5,769$). Classical Test Theory, Item Response Theory,
and Differential Item Functioning analysis will be employed to
determine if FCI items are equally fair for men and women.  We
acknowledge that a binary view of gender in physics education is
at best a first-order model, simplifying a wide range of
sociocultural factors and nuanced gender identities into two
categories~\citep{traxler_enriching_2016}. Nonetheless, this model
has been the basis for reporting many score differences on
standardized instruments such as the FCI. This work will focus on
fairness for men and women; future research should examine
fairness for other marginalized groups.

We will explore two dimensions of fairness: item fairness and
test construction fairness. An item is defined as being ``fair''
if men and women of equal ability have the same chance
of answering the item correctly. An instrument is defined as having
``test construction fairness'' if the instrument and items within the instrument
have similar performance on test evaluation metrics for men and women. An evaluation of
fairness is a crucial step in the test development process.

In this investigation, we find that if gender had been considered
relevant during test validation, a number of FCI questions would
have been flagged as poorly constructed or biased. At least some of these items
are consistent across multiple samples and with reports of
unfair items found in other studies. Finally, if these
items are removed, as would be common psychometric practice
during test development, a substantial portion of the canonical
gender gap disappears in at least some samples.

In the remainder of this introduction we will summarize ``gender
gap'' findings for the FCI, note the most popular student-based
causes that have been proposed, and then describe our psychometric
framework for analyzing the instrument. This framework draws in
part on that of Jorion {\it et al.}~\citep{jorion_analytic_2015},
which maps a process for validating conceptual inventories, but
which we expand by incorporating item fairness as part of the
process.

\subsection{``Gender gap'' investigations of the FCI}

The FCI has been used for measuring student conceptual gains in
introductory mechanics for nearly 25 years. For more than half
that period, published studies have documented an apparent gender
difference in item responses, overall scores, and
instructional gains. Madsen, McKagan, and Sayre provide an
overview of the research into the ``gender gap'' in conceptual
instruments used in Physics Education Research (PER)
\citep{madsen_gender_2013}. On average, male students outperform
female students by $13\%$ on pretests and $12\%$ on posttests of
conceptual mechanics instruments, the FCI and the Force and Motion
Conceptual Evaluation \cite{fmce}. Men also outperformed women by
$8.5\%$ on posttests of electricity and magnetism instruments.
This effect is nearly universal with only one of the seventeen
studies showing a female
advantage on the posttest. 

Most of the studies reported in the Madsen, McKagan, and Sayre
review follow common educational research practice which locates
the source of the gap within the students. Suggested influences in
gender-based performance differences include documented
differences in male/female high school physics class election
\cite{edu2009,nces2015,ets2016} and the effect of these
differences on college physics grades
\cite{sadler2001,hazari2007}. A large body of research also shows
differences in academic course grades
\cite{voyer2014gender,ets1997} and performance on cognitive tests
\cite{maeda2013,halpern,hyde1988gender,hyde1990gender} with women
scoring higher on verbal reasoning and men scoring higher on
spatial reasoning. Physics-specific variations on this research
have examined declared major, years of high school calculus, and
correlations with the Lawson test of scientific reasoning or other
standardized tests as a proxy for broader cognitive abilities (see
Madsen, McKagan, and Sayre, Table I for summary).

Many psychological factors have also been investigated to
explain gender differences such as mathematics anxiety
 \cite{else2010cross,ma1999meta}, science anxiety
 \cite{mallow1982,udo2004,mallow2010}, and stereotype
threat \cite{shapiro2012}. In physics education research, psychological
explanations have included self-efficacy, endorsement of gender
stereotypes, or attitudes toward physics \citep[][Table
I]{madsen_gender_2013}.

It is much harder to find studies that investigate gender bias in
university physics learning environments, though work in science
education has linked such bias to the greater attrition of women
from many STEM
fields~\citep{seymour_undergraduate_1992,seymour_loss_1995}.
Results show decreased gender gaps in classrooms using some active
learning curricula~\citep{lorenzo2006reducing,kost2009} which may
provide an avenue to reduce attrition. A great deal of work
remains to be done in this area, and it is likely to require
detailed qualitative data collection and analyses that are
substantially more time-consuming to conduct than pre/post conceptual
inventory measures.

A third possible source of conceptual inventory gender gaps, that
of bias in the test questions, can be analyzed by later
researchers even if it is not considered during instrument design.
For the FCI, several studies have highlighted items using
psychometric analysis that appear to function differently for
students of different genders. These findings have typically not
received as much attention as more student-centered explanations
for performance differences. The FCI continues to be used as a
diagnostic of student understanding, and in many cases to assign
course credit, despite a trail of evidence of gender bias. We will
highlight these studies in the following sections that expand on
the psychometric framework, and return to them in our discussion
of results.

\subsection{Validity framework}

Jorion {\it et al.}\ \cite{jorion_analytic_2015} described a
framework for evaluating the validity of conceptual inventories.
They applied this framework to multiple conceptual inventories
popular in engineering education and identified a number of flaws
with these instruments. Their framework collects some standard methods
of item analysis used in Classical Test Theory (CTT) and Item Response Theory (IRT)
\cite{crocker_introduction_1986}. Item analysis is usually performed at the
beginning of the test validation process to identify items which may be a threat
to the reliability of the instrument. An instrument with poor reliability cannot
have strong validity.
A review of the literature did not
identify any published work formally performing an item analysis for the FCI. The
Jorion framework will guide the item
analysis part of this work. The framework began by using
thresholds for item difficulty and discrimination from CTT
to flag potentially poorly-constructed items.
Items with poor performance on some psychometric measures will
be called ``problematic.''
The Item Characteristic
Curves from IRT were then examined to
determine if some items were problematic within the IRT model.
Cronbach's alpha and the inter-item correlations were then
calculated to identify items that may have reliability problems.
The factor structure of the instrument was then compared with the
factor structure published by the instrument's creators.

The framework does not address the issue of using a common instrument
for both pretest and posttest with student populations of varying
academic capability. Furthermore, it does not evaluate
item fairness, a critical oversight for conceptual
instruments used in class environments where some populations of
students are seriously underrepresented. We adopt the CTT and IRT measures used by
Jorion {\it et al.}, but will not consider factor analysis.
We extend the framework to include item fairness
analysis using Differential Item Functioning as discussed below.

The Jorion framework employs methods which should be performed at the
beginning of the process of producing reliable and valid
instruments. These methods are far from complete. Once a set
of reliable and fair items is identified, additional analysis is required to
demonstrate these items measure the intended constructs. There is an impressive
array of evidence attesting to both the face and criterion validity of the FCI as
well as its test-retest reliability for gender aggregated samples \cite{lasry2011puzzling,henderson2002common}.

\subsection{Classical Test Theory}

Establishing the validity of an instrument is a multifacted process that must be repeated
for all populations of interest.
A first step considers two basic tools of item analysis, difficulty and discrimination.
In CTT \cite{crocker_introduction_1986}, item difficulty, $P$, is
defined as the proportion of participants that answer an item
correctly for a given population (thus, higher values indicate
easier items). Item discrimination, $D$, is defined as \cite{crocker_introduction_1986}
\begin{equation}
D = P_{u}-P_{l},\label{eq:disc}
\end{equation}
where $P_{u}$ is the proportion of participants in the top 27\% of
the total score distribution answering the question correctly and
$P_{l}$ is the proportion of participants in the bottom 27\%
answering the item correctly. An item with low or negative discrimination
would be answered correctly by a substantial percentage of
low-scoring students and incorrectly by high scoring students, and
thus might indicate a question that was poorly phrased or mostly
answered by guessing.

For distractor-driven instruments, where the incorrect responses
are drawn from attractive alternate ideas, an item is judged to be
appropriate if its discrimination is above 0.2
\cite{jorion_analytic_2015,sadler_psychometric_1998,lindell_enhancing_2001}.
In addition, items should not be either too difficult or too easy,
resulting in difficulty cut-offs below 0.2 and above 0.8
\cite{jorion_analytic_2015}. Items that fall outside these cutoffs
will be classified as ``problematic'' and would normally be
considered for elimination during the test construction process.

While many studies employ the FCI, few report item level
statistics. Wang and Bao calculated CTT difficulty and
discrimination parameters for the FCI pretest of 2,800 students at
a large university in the US \cite{wang_analyzing_2010}. Five of
the items had difficulty parameters outside of the desired range
(items 1, 6, and 12 with $P>0.8$ and items 17 and 26 with $P<0.2$),
with none having discrimination less than $0.2$. Morris {\it et
al.} reported the item averages of 4,500 students pooling data
from multiple institutions and reported FCI items 5, 17, and 26
with $P<0.2$, but no items with $P>0.8$ \cite{morris2012item}.
Osborn Popp, Meltzer, and Megowan-Romanowicz reported FCI item level
scores for 4,775 high school students. For male students, items 1,
6 and 16 had $P>0.8$; for female students Item 26 had $P<0.2$
\cite{popp2011}.

CTT also provides measures of instrument reliability.
We use the most common of these, Cronbach's alpha or $\alpha$,
which is related to the average inter-item correlation.
Items in an instrument should be positively
correlated, so that if a student answers one item correctly it
implies that they are more likely to answer a second item
correctly.
Cronbach's alpha
generally increases as items are added to an instrument;
therefore, items for which $\alpha$ increases when an item is removed
are potentially problematic.

Lasry {\it et al.}\ assessed the overall reliability of the FCI
both by measuring test-retest performance and internal
consistency \cite{lasry2011puzzling}. Their study reported the
Kuder-Richardson reliability coefficient (KR-20), which had the
value $0.9$ for the initial application of the FCI and $0.865$
combining the initial test and a retest given one week later. The
KR-20 statistic is equivalent to Cronbach's alpha for dichotomous
items such as those used in the FCI. Values greater than 0.7
represent acceptable internal consistency \cite{nunnally}. Henderson
\cite{henderson2002common} also examined test-retest reliability
between the FCI as a graded posttest and as an ungraded quiz given
the following semester; excellent test-retest reliability was
measured in a sample of 500 university students. The FCI has also
been compared with an alternate test of conceptual knowledge of
mechanics, the FMCE; a high correlation of overall test scores,
$r=0.78$, was demonstrated \cite{thornton2009comparing}. As far as
we have been able to determine, no study has reported the change
in alpha as items are removed from the FCI.

One can also examine subscale reliability, whether subgroups of
questions thought to measure the same construct vary together.
Factor analysis is often
employed to
identify these subgroups. The FCI authors proposed a division of the instrument
into sub-categories \cite{fci}, but exploratory factor
analysis failed to reproduce this division
\cite{huffman_what_1995}. More recent analyses have resolved an
alternate factor structure \cite{scott2012exploratory}; however,
replication studies are needed to determine if these structures
are robust. Because there is not yet a consensus on the FCI factor
structure, we did not perform a confirmatory factor analysis.

\subsection{Item Response Theory}

CTT treats each item independently when calculating difficulty,
ignoring the repeated-measures nature of an
examination containing multiple items. CTT, therefore, ignores
correlations resulting from the differing abilities of test
takers. Item Response Theory (IRT) explicitly models the effect of
differing abilities by introducing a latent trait, $\theta_i$,
which varies by participant, $i$, and is related to the
probability that the participant answers a question correctly
independent of the item. IRT is an expansive topic with models for
many testing situations \cite{crcbook}. The model most closely
related to CTT is called the 2PL model, or 2-parameter logistic
model. This model assumes that each item, $j$,  has a
discrimination, $a_j$, and a difficulty, $b_j$. The probability,
$\pi_{ij}$, that participant $i$ answers item $j$ correctly is
given by the logistic function:

\begin{equation}
\label{eqn:2pl}
\pi_{ij}=\frac{\exp[a_j(\theta_i-b_j)]}{1+\exp[a_j(\theta_i-b_j)]}.
\end{equation}

Some authors rescale $a_j$ to map the logistic function
approximately onto the normal-ogive function. We will report the
untransformed discrimination $a_j$; that is, we will work in the
logistic metric as opposed to the normal metric. From 
(\ref{eqn:2pl}), the probability of any set of item responses can be
calculated and maximum likelihood estimation techniques employed
to fit the parameters $a_j$, $b_j$, and $\theta_i$.

An extension of the 2PL model that incorporates a third item
parameter designed to model random guessing behavior, the 3PL
model, has been used to investigate the FCI
\cite{wang_analyzing_2010}. The efficacy of the guessing parameter
for distractor-driven instruments such as the FCI has been
questioned \cite{morris2012item}. Morris {\it et al.} applied an
alternate method called ``Item Response Curve'' analysis which
replaces the ability, $\theta$, estimated by IRT with the overall
FCI score \cite{morris2006testing}. This method was compared with
Wang and Bao's IRT analysis \cite{morris2012item} and found very
small guessing parameters, indicating the 2PL model may be more
appropriate.

\subsection{Ability}
 IRT employs the term ``ability'' for the latent trait $\theta$
 which must be estimated to fit the IRT models. We will not
 report, nor make any use of $\theta$.
 In this work, when we refer to the ``ability'' of students, we
 mean their facility 
 at correctly answering conceptual
 physics questions like those contained in the FCI. No broader
 implications of general intellectual or academic capability are
 intended.

\subsection{Item Fairness}

The score on an evaluation instrument or item within that
instrument is fair to multiple groups of participants if members
of each group with the same ability generate similar outcomes.
Differential Item Functioning (DIF) analysis will be used to
explore whether the scores on individual FCI items are fair.

DIF analysis provides statistics to assess the score fairness of
items for subgroups of participants who have different abilities.
Many DIF statistics have been constructed; this work uses the
Mantel-Haenszel (MH) statistic \cite{holland1985alternate,
holland1988}, which is one of the most commonly used DIF measures
\cite{clauser1998}, and Lord's statistic, an IRT measure of DIF.
The MH statistic, $\alpha_{MH}$, is computed as a common odds
ratio for an item using the total score on the instrument to form
strata; thus, it pools the odds of a focal group (female students
in this study) to answer correctly compared to a reference group
(male students) for each level of ability, measured by overall
score. An effect size can be constructed through a logarithmic
transformation of the statistic
$\Delta\alpha_{MH}=-2.35\ln(\alpha_{MH})$ \cite{zwick1989}. This
effect size measure was adopted by the Educational Testing Service
(ETS) and is called the ETS Delta scale; it has been in use for
over 25 years \cite{zwick2012}. The ETS classifies
$|\Delta\alpha_{MH}|<1$ as negligible DIF,
$1\le|\Delta\alpha_{MH}|<1.5$ as small to moderate DIF, and
$|\Delta\alpha_{MH}|\ge 1.5$ as large DIF. Negative values
indicate an advantage to male students, positive values an
advantage to female students.

A substantial number of DIF statistics have been investigated for
IRT; we report Lord's statistic, $L$, which compares the
difference in difficulty parameters for the Rasch model with the
average difference in difficulty \cite{lord1980}. The Rasch model
is the 2PL model where the discrimination is constrained to one,
$a_i=1$. Multiplying by $2.35$ projects the statistic onto the ETS
Delta scale \cite{holland1985alternate,penfield2007handbook}.
\begin{equation}
\label{eqn:lord} L_{i}=2.35\cdot\bigg(b^F_i-b^M_i-\frac1n \sum_{j=1}^{n}
(b^F_j-b^M_j)\bigg),
\end{equation}
where $b^F_i$ is the female difficulty on item $i$, $b^M_i$ the
male difficulty, and $n=30$, the number of items. Lord's statistic
was selected because 
it corresponds to an effect size measure on
the Delta scale and it allows comparison with Osborn Popp, Meltzer,
and Megowan-Romanowicz's large study of DIF in high school
students \cite{popp2011}.

Dietz {\it et al.}\ used the MH statistic to evaluate DIF in an
approximately gender-balanced sample of 520 students and found FCI
items 4 and 9 were significantly biased against men and Item 23
biased against women ($p$s$<0.005$), all with large DIF
\cite{dietz_gender_2012}. They also present plots similar to
Figures \ref{fig:ark-combined} and \ref{fig:sample23-diff} (below,
Sec.\ \ref{sub:results-fair}). Their results showed many items
were substantially unfair to women; however, error bars were not
presented so it was difficult to assess whether these effects were
the result of sample variance. They acknowledge their results were
limited by sample size. While challenging to interpret because the
data was plotted on a logarithmic scale, if the averages remained
stable as sample size was increased, many items would exhibit
moderate to large DIF including items 6, 12, 14, and 27, which
will be identified as problematic in this study.

Osborn Popp, Meltzer, and Megowan-Romanowicz investigated DIF in
the FCI in a sample of 4,775 high school students who had completed
a high school physics course using Modeling Instruction
\cite{popp2011}. They used IRT with the Rasch model  and a DIF
statistic computed as the difference in the $b$ difficulty
parameter between men and women \cite{lord1980}. Their population
had an average difference in difficulty parameters of
approximately zero, and as such, their difference statistic was
equivalent to Lord's statistic, $L$, before multiplying by 2.35.
They found 14 items with significant DIF ($p<0.0017$) where a
Bonferroni correction had been applied to correct the $p$ value
for the number of statistical tests performed. Their statistic can
be converted to the ETS Delta scale by multiplying by 2.35. With
this conversion, for the significant items, Item 23 had large DIF
while items 4, 6, 9, 14, 15, and 29 had small to moderate DIF.

McCullough and Meltzer \cite{mccullough_differences_2001} compared
the performance of 222 algebra-based physics students on the
original FCI and a version where each problem was modified to have
a context thought to be more stereotypically familiar to women.
They found significant differences in performance on items 14, 22,
23, and 29. Using a similar methodology applied to non-physics
students, McCullough
\cite{mccullough_gender_2004} showed female performance did not
change while male performance decreased on the FCI modified to
stereotypically female contexts.

As such, there is substantial but inconsistent support for the
existence of gender unfair items in the FCI. This study seeks to
answer the following research questions: 
\begin{itemize}
\item RQ1: Are there FCI items
with difficulty, discrimination, or reliability values that would be identified as
problematic within CTT or IRT? If so, are the problematic items
consistent for male and female students?
\item RQ2: Are there FCI items where the CTT or IRT difficulty is substantially different for
male and female students?
\item RQ3: Are there FCI items which DIF
analysis identifies as substantially unfair to men or women?
\item RQ4: Are unfair FCI items identified by item analysis?
\item RQ5: Can differences in
answering by men and women for problematic items be explained by
an underlying physical principle or misconception?
\item RQ6: If small to moderate and large effect DIF items are removed from the FCI,
how does the gender gap change? 
\end{itemize}

\section{Methods}
\label{sec:methods}

\subsection{Samples}

This study will employ three datasets collected at four
US universities.
The FCI was revised after its initial
publication; this work uses the revised instrument published with
Mazur \cite{mazur1996peer} and available at PhysPort
\cite{physport}.
Racial/ethnic demographics were not available for individual students in the data but are reported at the university level.

{\bf Sample 1:} Sample 1 was collected from a large, southern
land-grant university enrolling approximately 25,000 students.
In 2012, university demographics by race/ethnicity were 79\% white, 5\% African American, 6\% Hispanic, with other groups each 3\% or less of the undergraduate population.
It had a
Carnegie classification of ``Highest Research Activity'' (or its
precursor, ``R1'') for the entire period studied. While strongly a
research university, its program was substantially lower rated (US
News Graduate Physics Ranking between 100--120 \cite{usnews})
than many of the institutions where PER has been
conducted. As such, this sample may contain students with lower
levels of preparation and somewhat lower academic capability than
those in which the gender gap has previously been explored. The
sample was collected from the spring 2002 semester to the fall
2012 semester. The dataset contains 4,509 complete pretest
responses (22.8\% female) and 4,716 complete posttest responses
(23.1\% female).

The FCI was applied as a pretest and posttest in the introductory calculus-based
mechanics class taken by scientists and engineers. Students received credit for
a good faith effort on the pretest and received a grade on the posttest. The course
was presented in the same format over the period studied and was overseen by
the same lead instructor for all semesters studied. This instructor created all
course materials including tests and homework assignments
and was the lead lecturer for approximately $75\%$ of the semesters
studied. For the other semesters, a graduate student or visiting instructor familiar with
the course delivered the lecture from the overall lead's notes. The course was
presented with two 50-minute lectures and two 2-hour laboratory sessions each week.
The lecture and laboratory components were tightly integrated. The lecture was traditional
while the laboratory featured a combination of research-based methods including small
group problem solving, hands-on open inquiry, and TA led demonstrations, as well as
traditional experiments. The course produced strong conceptual learning gains (Table \ref{tab:ave}).
Because of the stability of oversight, this sample does not
contain some of the confounding factors such as varying instructors bringing different
coverage and class policy that might be present in other large datasets.

{\bf Sample 2:} Sample 2 was drawn from two large, urban public
universities in the midwestern United States with similar student
profiles (primarily regional commuter students with a moderate
range of admission test scores).
In 2014-2015, the first university in the sample had racial/ethnic demographics of 71\% white, 13\% African American, 7\% international, with other groups 4\% or less.
The second university was 72\% white, 10\% African American, 6\% Hispanic/Latino, other groups 4\% or less.
The combined data contained 901
complete pretest responses (23.5\% female) and 649 complete
posttest responses (25.3\% female). This sample includes data from
fall 2014 to spring 2016 from several instructors. Instructional
styles ranged from traditional lecture, to moderately interactive
lectures using Peer Instruction \cite{mazur1996peer}, to heavily
interactive classes using Peer Instruction, Just-in-Time Teaching
\cite{novak_just-time_1999}, and cooperative group
problem-solving. Like Sample 1, students at these institutions
represent a more academically mixed sample than at many
PER-developer institutions. Neither institution held a Carnegie
classification of ``Highest Research Activity'' for the period
studied nor were ranked inside the top 150 physics graduate
programs by US News \cite{usnews}.

{\bf Sample 3:} Sample 3 was collected from a large, eastern
land-grant university enrolling approximately 30,000 students in
the spring 2015 semester. 
In 2015, the university's racial/ethnic demographics for undergraduates were 81\% white, 5\% African American, 6\% international, all other categories 4\% or less.
Data collection was part of an effort to produce cross-norming
data with an alternate mechanics conceptual evaluation routinely
given at the institution and to explore the effects of distractor
patterns on test performance \cite{twz}. Students received course
credit for a good faith effort. Minor modifications (reordering
the distractors) were applied to the FCI and found to have no
significant effect. The FCI was applied to both the introductory,
calculus-based mechanics and electricity and magnetism classes and
therefore this sample contains a longitudinal component; the
electricity and magnetism students had a larger time gap between
instruction and testing than the mechanics students. The dataset
contains 443 complete posttest responses 
(19\% female); pretest data
were not collected for Sample 3. This institution received the
Carnegie classification of ``Highest Research Activity'' in the
semester following the collection of the sample. While the physics
program had a similar US News ranking to that of Sample 1, other
university programs were not so highly ranked \cite{usnews}. This
institution also had a more open admission policy than that of
Sample 1 and, therefore, the Sample 3 student population should be less well
academically prepared than that of Sample 1.

\subsection{Measures and sample size}

This study reports results from CTT, IRT, and DIF analysis which
were discussed in the introduction.
Table \ref{tab:measures} summarizes the measures and their typical values.
For IRT analysis using the 2PL model, estimates of minimum
required sample size vary, with some authors suggesting that a
minimum of 200 is acceptable while others that samples of 500 are
required \cite{morizot2009toward}. While Sample 1 has sufficient
male and female students, there were too few female students in
samples 2 and 3 for accurate parameter estimation.

We tested the 3PL IRT model, which incorporates a guessing
parameter $c$, on Sample 1. It improved model fit somewhat, but
the guessing parameters extracted were much too large to be
credible for posttest results in a course producing the strong
conceptual performance of Sample 1. Similar departures from model
fit can be seen in the 3PL plots of Wang and
Bao~\cite{wang_analyzing_2010}, where many curves diverge from the
data at low ability. Additionally, uniform guessing parameter
models have been challenged for distractor-driven tests \cite{irt}. For these
reasons, this study will employ the 2PL model, which is also the
most closely related to CTT.

IRT introduces a model of student response patterns (Eqn. \ref{eqn:2pl}); the degree
to which this model fits the data was investigated.
IRT model fit can be evaluated for each item by dividing the
students into $G$ groups by their estimated ability, $\theta$, and
then estimating the goodness of fit between the predicted mean of
the group given by the 2PL model and the observed mean
\cite{bock1972estimating,reise1990comparison,yen1981using,wang_analyzing_2010}.
This produces a $\chi^2$ distributed statistic with $df=G-2$
degrees of freedom, because the 2PL model estimates two parameters
per item. Chi-squared tests have known problems with rejecting the
null hypothesis of good item fit for large samples
\cite{reise1990comparison}. To overcome this limitation, Cramer's
$V$ effect size statistic was also reported,
$V=\sqrt{\chi^2/(df\cdot N)}$, for both male and female students \cite{cramers}.
For $V$, $V=0.1$ represents a small
effect, $V=0.3$ a medium effect, and $V=0.5$ a large effect.
The number of groups used varies by
study; we selected a $G$ that ensured at least 100 students were
in each group, leading to $G=10$ for women and $G=33$ for men.
While some items were detected as significantly not fitting the
2PL model, no misfit represented even a small effect size.
Detection of some misfitting items was expected because of the
large sample size of the study. As such, the 2PL parameter
estimates should be accurate for this dataset. Item Characteristic
Curves were examined for all items. The plots for both male
students for all items and female students for most items had
similar visual fit to those presented in Wang and Bao
\cite{wang_analyzing_2010}. For all curves, significant misfit was
a result of variance between nearby bins and not an overall
failure of the 2PL model to fit the data.

DIF analysis with the MH
statistic groups students into strata. The finest grain 
possible
divides the students into groups with the same total test score;
less fine-grained strata can be formed by dividing students into
ranges of test scores. For example, Dietz {\it et al.}\ divided
students into five quantiles \cite{dietz_gender_2012}.
The large number of participants in Sample 1 allowed the
division by test score; five strata were employed for the smaller
Samples 2 and 3.
For Sample
1, both stratifications were compared and while
$\Delta\alpha_{MH}$ was somewhat different between the methods,
both yielded the same classification of DIF on the ETS Delta
scale.

All statistical calculations were performed using the ``R''
statistical software package \cite{R-software}. IRT
calculations were performed using the R package ``ltm''
\cite{ltm}, and
DIF calculations 
used the R package ``difR'' \citep{magis_general_2010}.

\begin{table*}
\caption{\label{tab:measures} Summary of item statistics, goodness-of-fit measures, and effect sizes reported in this study.}
\begin{ruledtabular}
\centering
\begin{tabular}{c|l|l}
Measure     & Description   & Usage and range notes \\ \hline
\multicolumn{3}{c}{CTT}\\\hline
$P$     & Item difficulty       & Values from 0 (hardest) to 1 (easiest); consider rejecting items with $P<0.2$ or $P>0.8$ \\
$D$         & Item discrimination   & Values from -1 (least discriminating) to 1 (most); consider rejecting items with $D<0.2$ \\
$\alpha$    & Cronbach's alpha  & Values in $[0,1]$; $\alpha>0.7$ indicates acceptable reliability \cite{nunnally}. \\
\multirow{2}{*}{$\phi$} & \multirow{2}{*}{Pearson correlation}  & Between items: Negatively correlated item pairs are potentially problematic. \\
        &               & Effect size of difference between $P_F$ and $P_M$: 0.1 small, 0.3 medium, 0.5 large \\ \hline
\multicolumn{3}{c}{IRT}\\ \hline
$b$     & Item difficulty       & Typical range of $-4$ (easiest) to 4 (hardest) \\
$a$         & Item discrimination   & Typical range of $-4$ (least discriminating) to 4 (most discriminating) \\
$d$     & Cohen's $d$   & Gender difference in calculated difficulty; 0.2 small, 0.5 medium, 0.8 large \\
$V$     & Cramer's $V$      & Goodness-of-fit; $0.1$ small misfit, $0.3$ medium, $0.5$ large \\ \hline
\multicolumn{3}{c}{DIF}\\ \hline
$\Delta\alpha_{MH}$ & ETS delta & $|\Delta\alpha_{MH}|<1$, negligible; $[1,1.5)$, small to moderate; $>1.5$, large \\
$L$     & Lord's statistic      & $|L|<1$, negligible; $[1,1.5)$, small to moderate; $>1.5$, large
\end{tabular}
\end{ruledtabular}
\end{table*}

This work reports the statistical significance of many
quantities and thus performs many statistical tests. To correct
for the inflation of Type I error rate, a Bonferroni correction
was applied to each set of analyses by dividing the critical
$p$ values by the number of tests performed. For example, for the
$\phi$ coefficient in Table \ref{tab:sum}, $p=0.05$ was changed to
$p=0.05/30=0.0017$ to correct for the 30 statistical tests
performed for the 30 FCI items.

\section{Results}
\label{sec:results}

Table \ref{tab:ave} presents overall FCI pretest and posttest
averages for the three samples. Significant gender differences
($p$s $<0.001$)
were measured for all applications of the FCI, with Cohen's $d$
\citep{cohen_power_1992} indicating small to medium effect sizes.
Cohen suggests that the practical
significance of a difference be considered as well as the effect
size \cite{cohen1990}. Some work suggests Cohen's initial effect
size criteria should be adjusted downward for educational research
\cite{hattie}. 
For Sample 1, course letter grades were available for about
two-thirds of the participants. For this subset, female students
($M=3.43$, $SD=0.75$) had somewhat higher grades measured on a
four-point scale than male students ($M=3.24$, $SD=0.89$) where
$M$ is the mean and $SD$ the standard deviation. While there is a
substantial literature showing superior female performance on
class grades \cite{voyer2014gender} 
and superior male performance
on standardized quantitative instruments
\cite{hyde1990gender,halpern}, this provides evidence that there
was not a substantial disparity between male and female academic
ability in Sample 1. The three samples present a spectrum of
course outcomes with Sample 1 generating the highest scores on the
FCI and Sample 2 the lowest. For Sample 1, female students closed
the pretest gender gap of 11\% somewhat to a posttest gap of 8\%,
while the gap changed little in Sample 2 from $12\%$ on the
pretest to $11\%$ on the posttest.

\begin{table}[htb]
\begin{ruledtabular}
\caption{\label{tab:ave} Pretest and posttest averages for all
samples. Mean ($M$) and standard deviation ($SD$) are reported as
percentages. No pretest was given in the Sample 3 classes. Cohen's
$d$ measures the effect size of the difference
between male and female scores.} \centering
\begin{tabular}{ l| c| c c| c c| c}
&&\multicolumn{2}{c|}{Male Students} & \multicolumn{2}{c|}{Female Students}&\\
&N&N& $(M \pm SD)\%$  &N&$(M \pm SD)\%$& $d$ \\\hline
\multicolumn{7}{c}{Sample 1}\\\hline
Pretest &4509 &3482 & $43\pm18$ & 1027   & $32\pm14$ &.69   \\
Posttest & 4716 & 3628&$73\pm17$&1088&$65\pm18$&.46\\\hline
\multicolumn{7}{c}{Sample 2}\\\hline
Pretest & 882 & 673 & $43\pm 20$ & 209   & $31\pm 15$ &.66    \\
Posttest & 610 & 464&$57\pm24$&146&$45\pm 18$&.56\\\hline
\multicolumn{7}{c}{Sample 3}\\\hline
Posttest & 443 & 361&$64\pm20$&82&$51\pm19$&.69\\
\end{tabular}
\end{ruledtabular}
\end{table}

\subsection{Difficulty and discrimination}

CTT and IRT were employed to examine the difficulty and
discrimination of the FCI. Item-level posttest results for Sample
1 are presented in Table \ref{tab:sum} and difficulty plotted in
Fig.\ \ref{fig:ark-combined}. The table presents the mean CTT
difficulty, $P$, CTT discrimination, $D$, IRT difficulty, $b$, and
IRT discrimination, $a$, for each FCI item. The CTT difficulties
for Sample 2 and 3 are plotted in Fig. \ref{fig:sample23-diff}.
Male and female students were investigated separately. The
standard deviations for the CTT parameters were calculated by
bootstrapping using 1000 sub-samples.
Table
\ref{tab:baddiff} presents the problematic items identified in the
FCI for each sample. Critically, many of the questions flagged for
female students in Table \ref{tab:baddiff} were not detected when
the data remained aggregated over gender.

\begin{table*}[htb]
\caption{\label{tab:sum} Classical Test Theory and Item Response
Theory results for Sample 1 for each FCI item. Male results are
marked (M) and female results (F). Significance levels have been
Bonferroni corrected for the number of statistics tests: ``a''
denotes $p<0.0017$, ``b'' $p<0.00033$, and ``c'' $p<0.000033$.}
\begin{ruledtabular}
\centering
\begin{tabular}{ c | c c c c c | c c c c c c c | c c}
&\multicolumn{5}{ c |}{Classical Test Theory} & \multicolumn{7}{ c |}{Item Response Theory} & \multicolumn{2}{c}{DIF}\\\hline
$\#$&  $P_M$         & $P_F$                  & $D_M$          & $D_F$           &$\phi$              &$b_M$                & $b_F$                &$a_M$               &$a_F$                 &$d$            &$V_M$         &$V_F$    &$\Delta\alpha_{MH}$        &\textit{$L$}   \\\hline
1   & .97$\pm$.00    & .95$\pm$.01    & .10$\pm$.01    & .13$\pm$.02     &$.04$       &-2.71$\pm$.16    &-2.78$\pm$.32     &1.63$\pm$.14    &1.30$\pm$.21   &.01               &.02                &.03                  & .33                           &.10                \\
2   & .66$\pm$.01    & .60$\pm$.02    & .56$\pm$.02    & .44$\pm$.04     &$.05^b$   &-.74$\pm$ .05     & -.61$\pm$.11     &1.09$\pm$.06    &.73$\pm$.08      &.05              &.02                &.03                  & .44                           &.50                  \\
3   & .91$\pm$.00    & .90$\pm$.01    & .22$\pm$.01    & .25$\pm$.03     &$.01$       &-2.15$\pm$.10    &-1.77$\pm$.12    &1.42$\pm$.09    &1.89$\pm$.20     &.07              &.02                &.04                  & 1.17$^b$                 &.84                   \\
4   & .62$\pm$.01    & .62$\pm$.01    & .59$\pm$.02    & .57$\pm$.03     &$.00$       & -.57$\pm$.04     & -.54$\pm$.07     &1.05$\pm$.05    &1.19$\pm$.11     &.01               &.02               &.05                  & 1.28$^c$                  &1.26$^c$                  \\
5   & .58$\pm$.01    & .50$\pm$.01    & .63$\pm$.02    & .65$\pm$.03     &$.07^c$   & -.35$\pm$.04     & -.03$\pm$.07     &1.24$\pm$.06    &1.08$\pm$.10     &.15$^c$       &.02$^a$        &.06$^a$           & .50                          &.35                    \\
6   & .91$\pm$.00    & .80$\pm$.01    & .22$\pm$.01    & .34$\pm$.03     &$.15^c$   &-2.34$\pm$.12    &-2.07$\pm$.25     &1.23$\pm$.09    & .75$\pm$.10     &.04                &.02                &.03                   & -1.43$^c$               &-1.34$^c$                      \\
7   & .88$\pm$.01    & .81$\pm$.01    & .22$\pm$.02    & .28$\pm$.03     &$.08^c$   &-2.69$\pm$.19    &-2.64$\pm$.40     & .81$\pm$.07     & .58$\pm$.10     &.00                &.02                &.03                   & -.45                        &-.21                     \\
8   & .89$\pm$.01    & .84$\pm$.01    & .26$\pm$.01    & .37$\pm$.03     &$.08^c$   &-2.13$\pm$.11    &-1.44$\pm$.10     &1.26$\pm$.08    &1.62$\pm$.16    &.12$^c$        &.02                &.06$^a$          & -.14                         &-.17                      \\
9   & .80$\pm$.01    & .84$\pm$.01    & .38$\pm$.02    & .40$\pm$.03     &$.03$       &-1.56$\pm$.08    &-1.44$\pm$.10     &1.12$\pm$.06    &1.59$\pm$.15     &.03                &.03$^c$        &.06$^a$           & 1.89$^c$                 &1.76$^c$                    \\
10  & .93$\pm$.00    & .90$\pm$.01    & .21$\pm$.01    & .28$\pm$.03    &$.05^a$   &-1.99$\pm$.08    &-1.72$\pm$.11     &1.95$\pm$.13    &1.92$\pm$.21     &.06                &.02                &.05                  & .39                          &.08                    \\
11  & .76$\pm$.01    & .73$\pm$.01    & .53$\pm$.02    & .63$\pm$.03    &$.02$       &-1.05$\pm$.04    & -.82$\pm$.06      &1.53$\pm$.07    &2.15$\pm$.18      &.09$^a$       &.03$^c$        &.06$^b$           & 1.31$^c$                 &.87$^b$                     \\
12  & .93$\pm$.00    & .80$\pm$.01    & .16$\pm$.01    & .31$\pm$.03    &$.17^c$   &-3.06$\pm$.22    &-2.16$\pm$.27     &.94$\pm$.08      & .71$\pm$.10       &.07                &.02                 &.02                 & -1.97$^c$                &-1.84$^c$                     \\
13  & .83$\pm$.01    & .79$\pm$.01    & .50$\pm$.02    & .57$\pm$.03    &$.04$       &-1.16$\pm$.04    & -.99$\pm$.06      &2.39$\pm$.12    &2.51$\pm$.22      &.08                 &.02                 &.05$^a$          & 1.22$^c$                 &.53                    \\
14  & .67$\pm$.01    & .40$\pm$.01    & .46$\pm$.02    & .44$\pm$.04    &$.23^c$   &-1.01$\pm$.07    & .63$\pm$.12       &.78$\pm$.05      & .66$\pm$.08       &.39$^c$         &.02$^a$         &.06$^b$           & -1.97$^c$               &-1.84$^c$                    \\
15  & .60$\pm$.01    & .66$\pm$.02    & .45$\pm$.02    & .54$\pm$.04    &$.05^b$   & -.64$\pm$.06     & -.71$\pm$.07     & .72$\pm$.05     &1.28$\pm$.11       &.02                 &.05$^c$         &.08$^c$          & 1.77$^c$                 &2.00$^c$                     \\
16  & .94$\pm$.00    & .91$\pm$.01    & .17$\pm$.01    & .28$\pm$.03    &$.04$       &-2.33$\pm$.11    &-1.71$\pm$.11     &1.51$\pm$.11    &2.15$\pm$.24       &.10$^b$        &.02                 &.05                 & .36                          &.17                     \\
17  & .55$\pm$.01    & .49$\pm$.02    & .67$\pm$.02    & .62$\pm$.03    &$.05^a$   & -.19$\pm$.03     & .03$\pm$.07       &1.42$\pm$.06    &1.19$\pm$.10       &.11$^a$        &.02                 &.05                 & .84$^c$                   &.62                      \\
18  & .57$\pm$.01    & .52$\pm$.02    & .68$\pm$.02    & .69$\pm$.03    &$.04$       & -.27$\pm$.03     & -.09$\pm$.06      &1.44$\pm$.06    &1.27$\pm$.11        &.09                &.02                &.05                  & 1.04$^c$                 &.70$^a$                      \\
19  & .87$\pm$.01    & .87$\pm$.01    & .29$\pm$.02    & .33$\pm$.03    &$.00$       &-1.86$\pm$.09    &-1.65$\pm$.12     &1.28$\pm$.08    &1.56$\pm$.16       &.04                 &.02                 &.06$^b$         & 1.35$^c$                  &1.14$^c$                     \\
20  & .65$\pm$.01    & .61$\pm$.01    & .53$\pm$.02    & .55$\pm$.03    &$.03$       & -.74$\pm$.05     & -.57$\pm$.09      &1.00$\pm$.05    & .95$\pm$.09       &.06                  &.02                 &.04                 & .75$^b$                   &.77$^b$                      \\
21  & .47$\pm$.01    & .23$\pm$.01    & .60$\pm$.02    & .29$\pm$.04    &$.20^c$   & .14$\pm$.04       &2.25$\pm$.33      & .99$\pm$.05     & .57$\pm$.08       &.38$^c$         &.04$^c$          &.05                & -1.86$^c$                 &-1.77$^c$                      \\
22  & .58$\pm$.01    & .34$\pm$.01    & .60$\pm$.02    & .42$\pm$.04    &$.20^c$   & -.38$\pm$.04      &1.11$\pm$.16      &1.08$\pm$.05    & .64$\pm$.08       &.45$^c$         &.03$^c$          &.07$^c$         & -1.61$^c$                 &-1.56$^c$                      \\
23  & .77$\pm$.01    & .45$\pm$.02    & .45$\pm$.02    & .43$\pm$.04    &$.29^c$   &-1.31$\pm$.06     & .35$\pm$.13       &1.15$\pm$.06    & .55$\pm$.08       &.43$^c$         &.02                 &.03                 & -2.70$^c$                 &-2.71$^c$                    \\
24  & .92$\pm$.00    & .83$\pm$.01    & .20$\pm$.01    & .32$\pm$.03    &$.12^c$   &-2.38$\pm$.13     &-1.79$\pm$.16     &1.26$\pm$.09    &1.10$\pm$.12      &.08                 &.02                &.04                 & -.94$^b$                   &-.98$^b$                     \\
25  & .54$\pm$.01    & .46$\pm$.01    & .74$\pm$.02    & .66$\pm$.03    &$.07^c$   & -.17$\pm$.03      &.14$\pm$.06        &1.72$\pm$.08    &1.31$\pm$.11       &.17$^c$        &.03$^c$         &.06$^a$         & .70$^a$                    &.32                   \\
26  & .32$\pm$.01    & .23$\pm$.01    & .66$\pm$.02    & .51$\pm$.04    &$.09^c$   & .64$\pm$.04       &1.15$\pm$.09      &1.65$\pm$.08     &1.44$\pm$.13       &.22$^c$        &.03$^c$         &.05                 & .40                           &-.08                  \\
27  & .77$\pm$.01    & .53$\pm$.02    & .38$\pm$.02    & .37$\pm$.04    &$.22^c$   &-1.58$\pm$.09     &-.27$\pm$.15       & .86$\pm$.05     & .45$\pm$.07         &.24$^c$       &.02$^a$         &.06$^a$         & -1.87$^c$                 &-1.80$^c$                  \\
28  & .71$\pm$.01    & .66$\pm$.01    & .63$\pm$.02    & .62$\pm$.03    &$.05^a$   & -.83$\pm$.04      &-.65$\pm$.07      &1.50$\pm$.07     &1.37$\pm$.12        &.08               &.02$^a$         &.05                 & .83$^b$                    &.56                \\
29  & .83$\pm$.01    & .85$\pm$.01    & .09$\pm$.02    & .14$\pm$.03    &$.02$       &-18.4$\pm$10      &-5.24$\pm$1.47   & .09$\pm$.05     & .34$\pm$.10          &.02               &.03$^c$          &.04                 & .64                           &1.55$^c$                  \\
30  & .62$\pm$.01    & .53$\pm$.01    & .59$\pm$.02    & .55$\pm$.04    &$.08^c$   & -.52$\pm$.04      &-.16$\pm$.08      &1.24$\pm$.06     & .86$\pm$.09          &.15$^b$      &.02                &.03                 & .19                            &.18               

\end{tabular}
\end{ruledtabular}
\end{table*}

\begin{table}[htb]
\caption{\label{tab:baddiff} CTT problematic items with $P<0.2$,
$P>0.8$, or $D<0.2$. }
\begin{ruledtabular}
\centering
\begin{tabular}{l c l}
Gender & Pre/Post & Problematic Items \\ \hline
\multicolumn{3}{c}{Sample 1}\\\hline
\multirow{ 2}{*}{Female}  & Pre   & 5, 11, 13, 15, 17, 18, 25, 26, 28, 30  \\
        & Post  & 1, 3, 6, 7, 8, 9, 10, 12, 16, 19, 24, 29       \\ \hline
\multirow{ 2}{*}{Male}    & Pre   &5, 6, 17, 18, 25, 26         \\
        & Post  &1, 3, 6, 7, 8, 9, 10, 12, 13, 16, 19, 24, 29  \\\hline
\multirow{ 2}{*}{Overall}    & Pre   &  5, 11, 17, 18, 25, 26        \\
        & Post  & 1, 3, 6, 7, 8, 9, 10, 12, 13, 16, 19, 24, 29 \\\hline
\multicolumn{3}{c}{Sample 2}\\\hline
\multirow{ 2}{*}{Female}  & Pre   &2, 5, 11, 13, 17, 18, 20, 25, 26, 28, 30         \\
        & Post  &17, 26, 29                  \\ \hline
\multirow{ 2}{*}{Male}    & Pre   &5, 17, 18, 26                \\
        & Post  &6, 12                      \\ \hline
\multirow{ 2}{*}{Overall}    & Pre   & 5, 11, 13, 17, 18, 26                \\
        & Post & 12                       \\ \hline
\multicolumn{3}{c}{Sample 3}\\\hline Female  & Post  & 1, 4, 6, 29
\\ \hline Male    & Post  &   1, 4, 6, 7, 12, 16, 24           \\ \hline
Overall    & Post  &  1, 4, 6, 7, 12, 16, 24           \\
\end{tabular}
\end{ruledtabular}
\end{table}


For Sample 1, all problematic items in the pretest had $P<0.2$
while all problematic posttest items had $P>0.8$. In Sample 2,
all problematic pretest items had $P<0.2$ while problematic
posttest items for male students had $P>0.8$ and problematic
posttest items for female students had $P<0.2$ (items 17 and 26)
or $D<0.2$ (item 29). For Sample 3, all problematic items had
$P>0.8$.

Examination of the gender-disaggregated posttest results in Table
\ref{tab:baddiff} identify Item 6 as problematic in 5 of the 6
samples while items 1, 12, and 29 were problematic in 4 of the 6
samples. Items 5, 17, 18, and 26 were problematic in all
gender-disaggregated pretest samples. There was little additional
commonality between the items flagged as problematic across all
samples. The problematic items in the Sample 1 posttest all had
very high scores. If the data was aggregated, Item 12 was
identified as problematic in all posttest samples.

IRT results can also be used to identify problematic items. One
FCI item, Item 29, produced difficulty parameters indicating the
IRT model was a poor fit for that item. None of the FCI items
showed the dramatic departures from model fit including negative
discrimination parameters identified in some of the inventories
examined by Jorion {\it et al.} \cite{jorion_analytic_2015}. As
such, IRT supports the identification of Item 29 as problematic.

\subsection{Reliability}
\label{ssec:rel}

Cronbach's alpha provides a measure of the overall reliability of
an instrument. If alpha increases with the removal of an individual
item, that item detracts from the overall instrumental reliability
and should be a candidate for elimination. Only posttest results
were explored for this analysis. For Sample 1, the FCI was
reliable with $\alpha=0.84$ overall, male students $\alpha=0.84$,
and female students $\alpha=0.83$. For male students, dropping
Item 29 increased alpha, while there was no item that could be
removed to increase alpha for female students. For Sample 2,
overall $\alpha=0.90$ with $\alpha=0.91$ for men and $\alpha=0.81$
for women. For male and female students, there was no item whose
removal increased alpha. For Sample 3, overall $\alpha=0.86$: with
$\alpha=0.85$ for male students and $\alpha=0.82$ for female
students. Removing Item 15 increased the overall alpha for both
male and female students. These reliability values were consistent
with those reported in Lasry {\it et al.} \cite{lasry2011puzzling}
and show that the FCI has strong internal consistency across a
variety of instructional settings. Cronbach's alpha of $0.7$ is considered
acceptable reliability; alpha of $0.9$ is required for higher stakes tests \cite{nunnally}.

To further investigate reliability, the correlation coefficient
between items can be calculated. In general, if a student answers
one item on a test correctly, the probability of answering a
second item correctly should increase; item scores should be
positively correlated. Jorion {\it et al.}
\cite{jorion_analytic_2015} calculated tetrachoric correlations
which assume the dichotomous variable, whether the question was
correct or incorrect, was derived from an underlying normal
continuum. This assumption seems unnatural for multiple-choice
physics questions where the student must either answer completely
correctly or incorrectly. Instead, we will report the Pearson
correlation which for two dichotomous variables is the $\phi$
coefficient~\cite{crocker_introduction_1986}. Tetrachoric
correlations were also calculated and in all cases had absolute
values greater than $|\phi|$. The significantly negatively
correlated ($p<0.05$) item pairs in Sample 1 were: male students,
$\{23,29\}$ and $\{29,30\}$ and female students $\{8,21\}$, $\{15,27\}$,
and $\{29,30\}$. In Sample 2, there were no significantly negatively
correlated item pairs for male students; for female students, only
items $\{12,29\}$ were significantly negatively correlated. For
Sample 3, no question pairs were negatively correlated for men,
while $\{7,15\}$ and $\{9,12\}$ were significantly negatively
correlated for women.  Both the correlation analysis and Cronbach's alpha support the identification of Item 29 as
problematic. Many of the items which were negatively correlated
will be identified as unfair in DIF analysis: items 9, 12, 15, 21,
23, and 27.

\subsection{\label{sub:results-fair}Item Fairness}

An item  is ``fair'' if students of the same ability from
two populations produce equal scores on the item.
Item fairness will first be investigated under the assumption
that male and female students are of equal abilities, then DIF
analysis will be applied to explore fairness without the
assumption of equal abilities. For this analysis, Samples 2 and 3
contain an insufficient number of female students to draw strong
statistical conclusions. The results of these samples will be examined
only in reference to Sample 1.

This work uses the terms ``ability'' and ``fairness,'' which are
common within the test development literature. Both terms have
broad colloquial meanings outside this literature, and as such, it
is important that the reader interpret these terms by their narrow
meaning. Ability is used to mean only the proficiency with which
students answer test items---in this case, conceptual physics
problems on the FCI. Fairness analysis depends on the
assumptions made about ability. If two groups have the same
proficiency in conceptual physics, then items where the groups
score differently do not test the two groups in the same way: the
items are unfair. If the assumption of equal proficiency is not
true, then items can score differently because of the differences
in the groups and a difference in score does not imply an unfair
problem. DIF analysis does not assume the two groups have equal
proficiency in conceptual physics, but uses the score on the FCI
as a measure of proficiency. In DIF analysis, an item is unfair if
the two groups have a larger difference in score than one would
predict from the difference in overall test score.

\subsubsection{Equal Ability Analysis}

\begin{figure*}[htb]
\includegraphics[width=\textwidth]{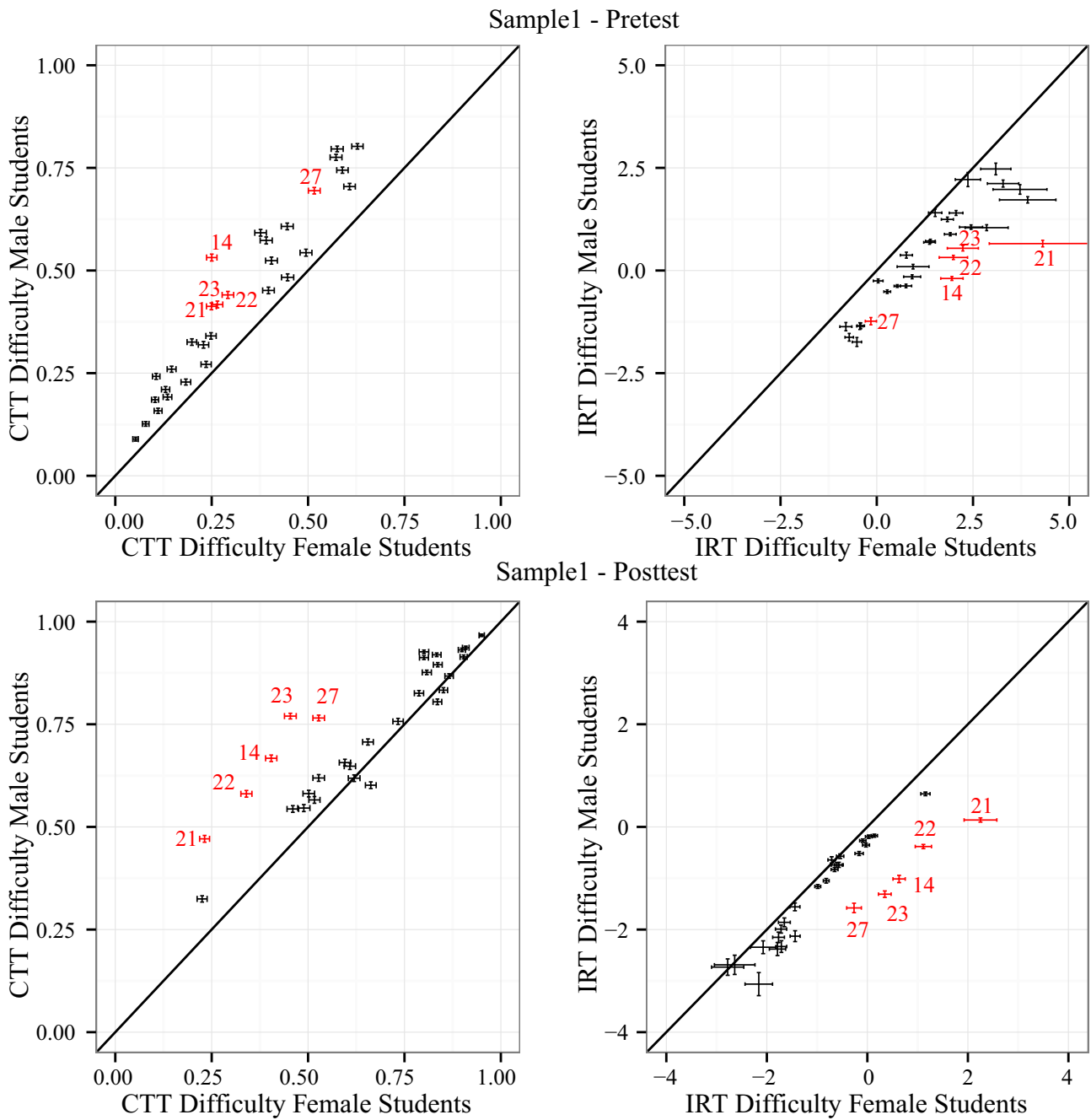}
\caption{\label{fig:ark-combined} CTT and IRT results for Sample
1. Items 14, 21, 22, 23, and 27 are marked in red and labeled. A line of slope one
is drawn to allow comparison of male and female difficulty. Error bars represent one standard deviation in each direction.}
\end{figure*}

If one assumes that male and female students have an equal ability
to answer conceptual physics questions correctly, then a fair FCI
item is one where the difficulty is equal for male and female
students. 
Under this assumption, which is supported by the higher course grades of female students, item
fairness can be explored by plotting the difficulty for
male students against the difficulty for female students. Figure
\ref{fig:ark-combined} shows this plot for the Sample 1 pretest
and posttest. A line of slope one is drawn on all plots; perfectly
fair questions would fall on this line (the fairness line). Items
unfair to women fall above the fairness line for the CTT plots and
below the line for IRT plots. For the posttest
scores, Fig.\ \ref{fig:ark-combined} has three striking features:
(1) most items are significantly unfair to women (the error bars
do not overlap the fairness line); (2) five items, 14, 21, 22, 23,
and 27, stand out as substantially unfair to women by falling well
off the fairness line; and (3) most other items fell fairly close,
but on the unfair to women side, of the posttest fairness line.
The substantially unfair items are plotted in red and
numbered in the figure. Similar plots were explored for item
discrimination and did not show any pattern of item bias. We will
focus on item difficulty for the remainder of the study.

To determine if the differences in performance in the CTT plot in
Fig.\ \ref{fig:ark-combined} were statistically significant and to
estimate effect sizes, the phi coefficient, $\phi$, was calculated
for each item and is included in Table \ref{tab:sum}. The
significance values for $\phi$ were calculated using the
chi-squared 
test of independence on the two-by-two
table of male and female correct and incorrect answers for each
problem. The $\phi$ coefficient is related to $\chi^2$ by
$\phi=\sqrt{\chi^2/N}$ where $N$ is the number of students. The
$\phi$ coefficient is equivalent to the two-point Pearson
correlation coefficient for dichotomously scored items and
provides a measure of effect size (Table \ref{tab:measures}).
For many items, male and female scores were
significantly different. For items 6, 12, 14, 21, 22, 23, 24 and
27, male and female difficulty scores were significantly different with a
small effect size. This set of items contains most of the items
which will be identified as significantly unfair by DIF analysis. The $\phi$
coefficient above is mathematically similar to the $\phi$ coefficient
in Sec.\ \ref{ssec:rel}; however, their use is conceptually different. In Sec.\ \ref{ssec:rel},
$\phi$ is used as a measure of association, so large $\phi$ indicates strongly
correlated items. In this section, $\phi$ is used as a measure of independence
and large $\phi$ indicates that the item difficulty is different for
men and women (small $\phi$ indicates the difficulty is independent of gender).

A similar analysis was used to explore whether differences in the
IRT difficulty coefficients were
significant.
Table \ref{tab:sum} shows these results in the Cohen's $d$ column.
The results were similar to those using the CTT
difficulty; the gender difference in items 14, 21, 22, 23, 26 and
27 was significant ($p$s$<0.001$) with a small to medium effect
size.

One item, Item 29, produced difficulty and discrimination
parameters that suggest the underlying IRT model was a poor
approximation for this item. The model was re-fit removing this
item. Parameter estimates changed very little; as such, the values
for the original model including Item 29 are reported.

The FCI pretest results for Sample 1 are also presented in  Fig.\
\ref{fig:ark-combined}. The five substantially unfair questions
identified in the posttest (14, 21, 22, 23, 27) were among the
most unfair questions in the pretest plots; however, many
additional questions were also substantially more difficult for
women. The IRT variance for women was also substantially higher
than in the posttest. Many pretest differences 
were reduced by instruction and many questions moved
substantially closer to the fairness line in the posttest, except
items 14, 21, 22, 23, and 27. As such, instruction reduced much of
the gender gap in the pretest questions but failed to address the
gap for a subset of the questions.

\begin{figure*}[htb]
\includegraphics[width=\textwidth]{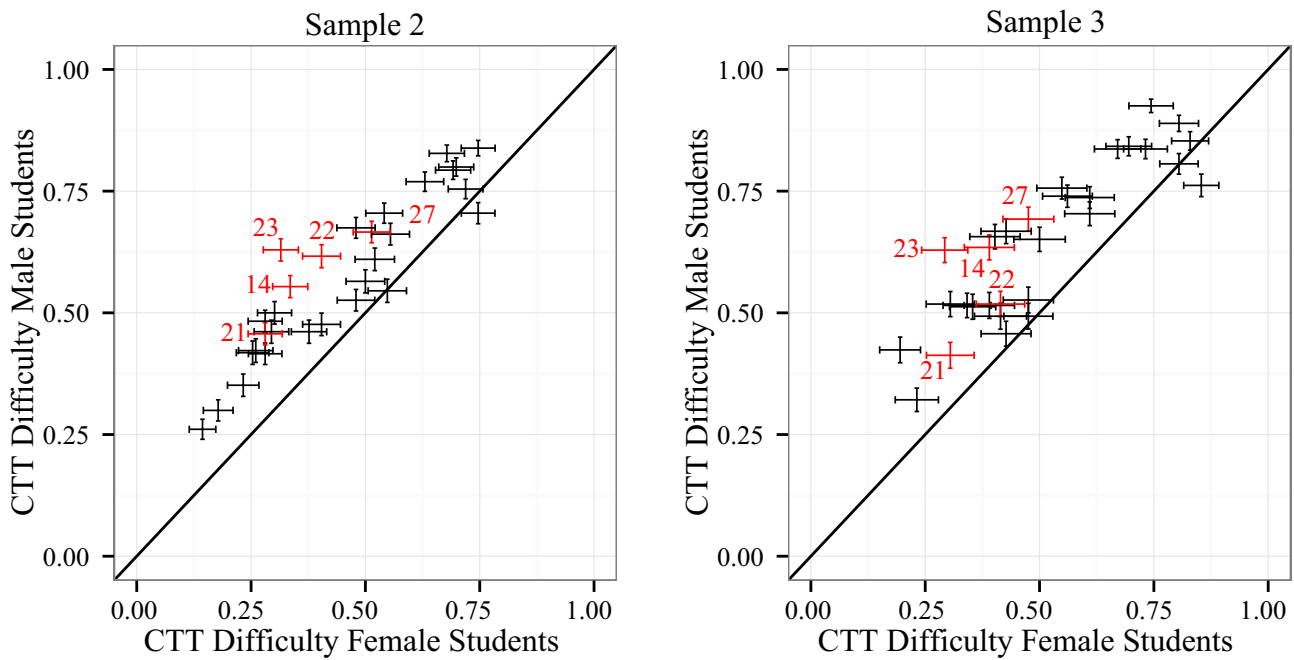}
\caption{\label{fig:sample23-diff} CTT posttest difficulty results for Sample
2 and 3. Items 14, 21, 22, 23, and 27 are marked in red. A line of slope one
is drawn to allow comparison of male and female difficulty. }
\end{figure*}

\begin{figure*}[htb]
\includegraphics[width=\textwidth]{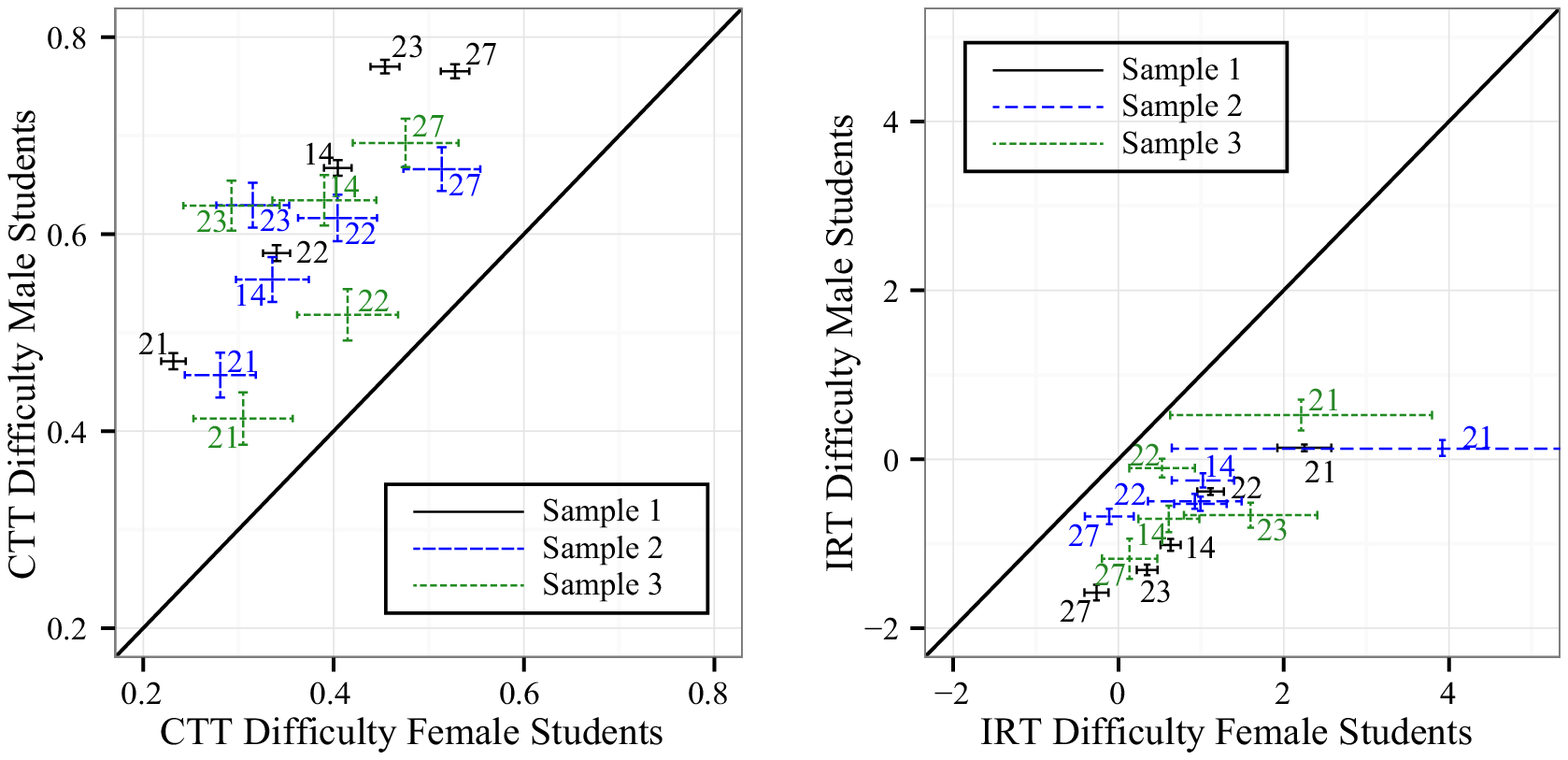}
\caption{\label{fig:combined} CTT and IRT  posttest difficulty
scores for male and female students for problematic items from all
samples. A line of slope one is drawn to allow comparison of male
and female difficulty. The item number for each problem is also
labeled. The IRT difficulty of Sample 2, Item 23 is not labeled;
the point overlays that of Sample 2, Item 22.}
\end{figure*}

Figure \ref{fig:sample23-diff} presents a plot of CTT posttest difficulty for Samples 2
and 3 with items 14, 21, 22, 23, and 27 also colored in red and labeled. The
much smaller sample size  caused the error bars of
many points to overlap, but many of the five most problematic items in
Sample 1 were also at the outside of the item envelope in Samples
2 and 3. 
Samples 2 and 3 
were too small for reliable IRT parameter
estimation. Figure
\ref{fig:combined} overlays plots of items 14, 21, 22, 23, and 27 for all
samples; the similarities, particularly in the CTT plot, are quite
strong. This supports the identification of these five questions
as generally unfair, not simply unfair because of some artifact of
either student population or instruction in Sample 1.
IRT results for Samples 2 and 3 are included in Fig.\ \ref{fig:combined},
but should be interpreted with caution.

This analysis was performed assuming men and women have equal
facility in answering conceptual physics problems. This assumption
may be less accurate on the pretest, where the gender gap was
larger. DIF analysis, which does not assume equal ability, will
identify fewer unfair pretest items.

\subsubsection{Differential Item Functioning Analysis}

The analysis of the previous section compared male and female
students and found significant differences in difficulty for many
FCI items under the assumption of equal male and female ability.
The clustering of many items
near the fairness line in Fig.\ \ref{fig:ark-combined} suggests
that, while there may be some overall difference in conceptual
performance between men and women, most items were only somewhat more difficult for women than men.

DIF analysis relaxes the assumption of equal ability and replaces
it with the assumption that the overall score on the instrument is
an accurate measure of ability. Table \ref{tab:sum} reports
$\Delta\alpha_{MH}$ for each item in Sample 1, stratified by total test score. 
Eight FCI
items demonstrated large DIF (9, 12, 14, 15, 21, 22, 23, 27), where
9 and 15 were biased in favor of female students.
This set includes most items identified
as significantly biased with a small effect size in the previous section.
 Seven additional questions demonstrated small to moderate DIF.

DIF analysis can also be carried out using the results of IRT.
We used Lord's statistic, $L$, which is mapped to the same range as $\Delta\alpha_{MH}$ and
reported in Table \ref{tab:sum}. The Lord's statistic results agreed with the high DIF
classification provided by $\Delta\alpha_{MH}$ except that Item 29 was also
flagged as high DIF favoring women. The small to moderate DIF results were
less consistent, and the two statistics disagreed on items 3, 11, 13, and 18. None of
these four items were ultimately identified as biased in the reduced FCI instrument
constructed to answer RQ6. This provides evidence of the efficacy of employing
both CTT and IRT analysis to complement one another. Note, Lord's
statistic is calculated with the difficulty parameters in the Rasch model
in which the discrimination is set to one ($a_i=1$). The difficulty parameters reported
in Table \ref{tab:sum} are for the 2PL model; therefore, Eqn.\ \ref{eqn:lord} cannot be used
to compute Lord's statistic using $b_i$ from Table \ref{tab:sum}.
Further, the difficulty parameter calculated in the Rasch model for Item 29
was reasonable allowing $L$ to be calculated for this item even through
the difficulty and discrimination in the 2PL model were problematic.

DIF analysis was also attempted for Sample 2 and 3 by stratifying
students into five quantiles to reproduce the analysis of Dietz
{\it et al.} \cite{dietz_gender_2012}. The stratification into 5
quantiles left only a few women in the highest scoring quantile
and the results were strongly dependent on the number of quantiles
selected. We concluded that the number of female students in
samples 2 and 3 was insufficient for accurate DIF analysis.

DIF analysis was also performed on the Sample 1 pretest. With the
much larger variance seen in Fig.\ \ref{fig:ark-combined} and the
generally weaker pretest performance of women, few items were
detected as significantly biased. The DIF results for the FCI
pretest for Sample 1 detected only Item 14 as having large DIF;
items 4, 12, 19, and 26 demonstrated small to moderate DIF. This
difference between pretest and posttest is consistent with the
observation that women close the score gap with men on many
problems post instruction. Because DIF stratifies by overall test
score, a smaller gap can be considered unfair on the posttest than the pretest
if the overall posttest gap is smaller than the pretest gap.

\subsection{Item-level Analysis}

The distribution of student answers for the 5 most unfair items of
Sample 1 are shown in Table \ref{tab:item}. Female students
preferentially selected one of the distractors for each item. For
Samples 2 and 3, the selection of distractors was less uniform,
possibly because of the relatively small number of female students
in Samples 2 and 3 or because of the lower overall FCI scores for
these samples. The differences in responses observed between male
and female students in Sample 1 may have resulted from one or more
physics concepts that were not mastered by female students or from
surface features of the problem's context that made the problem
more difficult for female students. Examination of these problems
does not immediately suggest a common physics concept underlying
the incorrect answers.

\begin{table}[htb]
\caption{\label{tab:item} Answer distribution for problems with
large gender differences in CTT and IRT difficulty in Sample 1.
Correct answers are bolded.}
\begin{ruledtabular}
\centering
\begin{tabular}{ c |c | c c c c c}
&&\multicolumn{5}{ c }{Response}\\\hline
\#&Gender&(a)&(b)&(c)&(d)&(e)\\\hline
\multirow{2}{*}{14}& Male & 10\% & 4\%&18\%&{\bf 67\%}&0\%\\
&Female&30\%&12\%&17\%&{\bf 40\%}&0\%\\\hline
\multirow{2}{*}{21}&Male&2\%&5\%&39\%&7\%&{\bf 47\%}\\
&Female&3\%&16\%&53\%&5\%&{\bf 23\%}\\\hline
\multirow{2}{*}{22}&Male&31\%&{\bf 58\%}&1\%&9\%&0\%\\
&Female&55\%&{\bf 34\%}&1\%&9\%&0\%\\\hline
\multirow{2}{*}{23}&Male&7\%&{\bf 77\%}&6\%&8\%&1\%\\
&Female&25\%&{\bf 45\%}&13\%&14\%&2\%\\\hline
\multirow{2}{*}{27}&Male&19\%&3\%&{\bf 77\%}&1\%&0\%\\
&Female&40\%&6\%&{\bf 53\%}&1\%&0\%\\
\end{tabular}
\end{ruledtabular}
\end{table}

For Item 14 (bowling ball falling out of an airplane), the most
popular distractor for female students was the rearward parabolic
trajectory, while the most popular distractor for male students
was a linear forward trajectory. Item group 21--24 concerns a
scenario where a sideways-drifting rocket turns on its engine for
a period and then off again. The differences in items 21 to 23
seemed to result from students answering the question correctly
for the assumption that the force was an impulse force. The preferentially
selected distractor for items 21 and 22, for both men and women,
was correct for an impulse force. The relatively random pattern of
incorrect answers on Item 23 (turning off the engine) might result
because the question does not make sense if one is assuming the
engine is already off. The question group does state that the
engine is on for the entirety of items 21 and 22. The text employs
the verb ``thrust''; colloquially, the verb ``to thrust'' means to
``push or drive quickly and forcibly'' \cite{thrust}. Item 27
concerns a large box being pushed across a horizontal floor, and
the preferred distractor across genders was that the box comes
immediately to a stop.

The problem contexts described above might be more familiar on
average to men through everyday experience (Item 27) or through
greater exposure to physically realistic video games and movies
(items 14, 21--23). However, it is difficult to construct such an
explanation that would not apply equally to items 9 and 15
(kicking a hockey puck and pushing a broken-down truck), which had
a large DIF favoring women. Wilson {\it et al.}\ showed that gender
differences in physics questions used in physics competitions were
particularly large for two-dimensional motion and projectile
motion problems \cite{wilson2016}. However, questions identified
in the current study as unfair to both men and women fall in these
categories. Without the identification of a physical principle or
common misconception that unifies the items, the determination of
the origin of the gender difference must be left for a future
study.

\subsection{An Unbiased Force Concept Inventory}

To construct an unbiased version of the FCI, items were
iteratively removed, $\Delta\alpha_{MH}$ recalculated, and
additional items removed until no item in the FCI showed small to
moderate or large DIF for Sample 1. This process removed the 8
questions with large DIF as well as items 6 and 24, producing a reduced
instrument containing FCI questions: 1, 2, 3, 4, 5, 7, 8, 10, 11,
13, 16, 17, 18, 19, 20, 25, 26, 28, 29, and 30. For Sample 1,
this 20-item instrument reduced the gender gap on the posttest to
$4.3\%$ from the original $8.0\%$ with men scoring $(73.1\pm19)\%$
and women scoring $(68.7\pm 19)\%$. The difference was still
significant [$t(1761)=6.55, p<0.001$] but with a substantially
smaller effect size, $d=0.23$. The total scores on the original
and reduced instruments were highly correlated for both male and
female students ($r=0.96$) where $r$ is the
Pearson correlation coefficient. If the instrument is further
reduced by removing Item 29 which was shown to be problematic in
item analysis and in DIF analysis with Lord's statistic,
the gender gap increases slightly to $4.7\%$. The reduced instrument still
contains a number of items originally calculated to have small to moderate DIF (Table \ref{tab:sum}).
The DIF of these items became negligible after the higher DIF items were
removed.

For Samples 2 and 3, the reduced instrument did not substantially
reduce the gender gap. For Sample 2, the original gender gap of
$12.9\%$ became $11.4\%$ for the 20-item instrument and $12.2\%$
with the further removal of Item 29. For Sample 3, the original
gender gap of $13.5\%$ was reduced to $12.7\%$ for the 20-item
instrument, but increased to $13.8\%$ with the removal of Item 29.

The pretest gender gaps changed little on the reduced instrument. For Sample 1,
the gender gap on the 20-item FCI was $9.9\%$ which was somewhat smaller than
the gender gap of $11.9\%$ on the original 30-item FCI. Further removing Item 29
increased the gap to $10.1\%$. For Sample 2, the gender gap on the 20-item instrument
was $10.3\%$ which was somewhat smaller than
the gender gap of $12.3\%$ on the original 30-item FCI. Further removing Item 29
increased the gap to $10.6\%$.

\section{Discussion}

This study sought to answer six research questions; these will be
addressed in the order proposed. We then consider larger patterns in
prior research in light of our results.

\subsection{Research questions}

{\it RQ1: Are there FCI items with difficulty, discrimination, or
reliability values that would be identified as problematic within
CTT or IRT? If so, are the problematic items consistent for male
and female students?} CTT identified few areas where the FCI or
items within the FCI were uniformly problematic across all
samples. Aggregating men and women, Item 12 was flagged as
problematic in all posttest samples. Items 5, 11, 17, 18, and 26
were identified as problematic in both aggregated pretest samples.
Item 6 was problematic in 5 of the 6 gender-disaggregated posttest
samples. Items 1, 12, and 29 were identified as problematic in 4
of the 6 gender-disaggregated posttest samples. Items 5, 17, 18,
and 26 were identified as problematic in all gender-disaggregated
pretest samples. Identification of difficulty parameters outside
the desired range likely resulted from the application of the FCI
at multiple institutions with differing student populations as
both a pretest and posttest. This caused some items to be flagged
on the pretest with $P<0.2$ and on the posttest with $P>0.8$. IRT
and reliability analyses further supported the identification of
Item 29 as problematic.

The items and the number of items identified as problematic
differed between male and female students. More items were
problematic for female students in Sample 1 and Sample 2 on the
pretest. More items were problematic for male students in Sample 3
on the posttest. Crucially, an analysis that aggregated
men and women, the ``Overall'' rows in Table \ref{tab:baddiff},
would reach conclusions accurate for male students but often very
inaccurate for female students.

The problematic CTT and IRT  items provide less accurate
information about the knowledge of the student than
non-problematic items by either being too hard, too easy, or too
likely to answered correctly by weak students (or incorrectly by
strong students). Many items on the FCI provide less information
about female students than male students in the Sample 1 and 2
pretest; the FCI contains many items that provide  less
information about male students in the Sample 3 posttest.
While these problems almost certainly resulted from using one
instrument in multiple environments both as a pretest and
posttest, instructors should be aware that the FCI can provide
results with different levels of validity for different student
populations even in the same testing conditions. As such, its results
should used with caution for these populations.

{\it RQ2: Are there FCI posttest items where the difficulty is
substantially different for male and female students?} FCI items
6, 12, 14, 21, 22, 23, 24, and 27 in Sample 1 demonstrated a
significant gender bias in item difficulty (Table \ref{tab:sum})
in CTT with a small effect size. IRT identified items 14, 21, 22,
23, 26, and 27 as significantly unfair with a small effect size.
The interpretation of items 14, 21, 22, 23, and 27 as
substantially unfair was supported by graphical analysis of
Samples 2 and 3 (Fig.\ \ref{fig:combined}).

{\it RQ3: Are there FCI items which DIF analysis identifies as
substantially unfair to men or women?}
In Sample 1, DIF analysis confirmed the unfairness of items 12, 14, 21, 22, 23, and
27 and further identified items 9 and 15 as having large DIF;
items 9 and 15 were biased in favor of women. Iteratively removing
high DIF items also showed items 6 and 24 with
high DIF once the highly biased items were removed. Because DIF depends on
overall test score, the DIF of an item changes as unfairly
functioning items are removed from an instrument. Items
3, 4, 11, and 18 demonstrated small to moderate DIF; however, the DIF of these
items became negligible 
as the more unfair items were removed
to form the 20-item unbiased FCI.

The Sample 1 posttest results of this study were fairly consistent
with those of other work. The Sample 1 results of this study
supported the advantage for women in Item 9 found in Deitz {\it et
al.}\ \cite{dietz_gender_2012} (large DIF) and Osborn Popp {\it et
al.}\ \cite{popp2011} (small to moderate DIF). This study also
supported the large DIF toward men of Item 23 found in both Deitz
{\it et al.}\ and Osborn Popp {\it et al.} Deitz {\it et al.}\ did
not report small to moderate DIF items; however, from the graph
presented \citep[][Fig. 4]{dietz_gender_2012} it seems likely Item 15 would be found biased towards
women with items 12, 14, and 27 biased towards men, 
consistent with this work. The graph also suggests Item 30 may
also be biased toward men. Osborn Popp {\it et al.} also
identified items 4, 9, 15, and 29 with small to moderate DIF
toward women and and items 6 and 14 with small to moderate DIF
toward men. The current study identified Item 4 as unfair (small to moderate DIF) in
Sample 1, as was reported in Deitz {\it et al.} (large DIF) and
Osborn Popp {\it et al.} (small to moderate DIF); however, the DIF of this item
became negligible as more highly biased items were removed from the FCI. Items 14, 22,
23, and 29 were also identified by McCullough and Meltzer as
demonstrating significant differences between male and female
answering patterns when the context of the question was modified
to be more stereotypically female oriented
\cite{mccullough_differences_2001}.

Combining the results of this study with those of previous
research strongly identifies a set of biased items in the FCI. The
relatively consistent pattern of items 6, 9, 12, 14, 15, 22, 23,
and 27 being identified as gender biased in multiple studies
strongly indicates the use of these questions should be
reconsidered. This study additionally suggests that items 21 and 24 should
be reconsidered because of bias and Item 29 because of recurring reliability issues.
Removing all these items would produce a
19-item instrument. Because the FCI has not demonstrated a
consistent factor structure \citep{huffman_what_1995} and
therefore is primarily a single factor instrument measuring the
degree to which a student possesses a ``Newtonian Force Concept,''
a 19-item instrument should measure this facet with approximately
the same accuracy as a 30-item instrument.

{\it RQ4: Are unfair FCI items identified by item analysis?} Most items
ultimately identified as unfair in the FCI
were not uniformly flagged as problematic
by CTT or IRT item analysis.
Only items 6 and 12 were detected as problematic in both DIF and
item analysis using discrimination and difficulty cutoffs.  Item
fairness analysis is therefore a complementary method that
provides additional information beyond item analysis methods. CTT
and IRT difficulty, discrimination, and reliability checks do not
guarantee item score fairness. Some additional high DIF items were
identified in reliability analysis but only after disaggregating
by gender.

{\it RQ5: Can differences in answering by men and women for
problematic items be explained by an underlying physical principle
or misconception?} Examining answer patterns for the
biased questions in Sample 1 did not identify an underlying
physical principle or misconception that was shared by all or some
combination of the questions. This makes it unlikely a general
failure of instruction either by the course studied or within the
academic background of the students studied accounted for the
differences identified. Further experimental investigation such as
that performed by McCullough and Meltzer
\cite{mccullough_differences_2001} will be required to determine
the origin of the gender differences.

{\it RQ6: If small to moderate and large effect DIF items are
removed from the FCI, how does the gender gap change? } For Sample
1, removal of all questions with small to large DIF resulted in a
20-item instrument. The gender gap on the posttest using this
reduced instrument was $4.3\%$ ($d=0.23$) which was substantially
smaller than the original posttest gender gap of $8.0\%$
($d=0.46$) with half the effect size. Item fairness, then, does
not explain all the gender gap in the FCI but accounts for about
half of the gap in this sample. The gender gap on the 20-item
gender-neutral instrument's posttest would be the second smallest
FCI gap reported \cite{madsen_gender_2013}.

The reduced instrument did not significantly reduce the gender gap
in Samples 2 and 3. An explanation may be found by comparing Fig.\
\ref{fig:ark-combined} to Fig. \ref{fig:sample23-diff}. In Sample
1, female students improved on many items that were substantially
unfair in the pretest, leaving only a few items where women were
substantially off the fairness line on the posttest. Sample 2 and
3 students did not demonstrate the same degree of progress, and
women in these samples do not show a substantial number of nearly
fair questions post-instruction.

\subsection{Insights into previous studies}

Some studies have suggested that more interactive teaching methods
lower the gender gap
\cite{lorenzo2006reducing,pollock2008comparing,kohl2009introductory};
however, this effect has not been consistently reproduced
\cite{pollock2007reducing}. Some research-based instructional
methods were employed in the lecture portions of Sample 2 and 3
while Sample 1 combined a traditional lecture with an interactive,
inquiry-based laboratory experience. While the courses from which
all three samples were drawn presented some interactive or
research-based instruction, the primary differences between the
courses seems to be the overall conceptual learning outcome
measured by FCI posttest scores. Excluding the items showing
substantial gender bias, the course measured in Sample 1 produced
posttest results where the performance of male and female students
were more similar (most results fell near the fairness line). The
posttest results for Sample 2 and 3 have many more items
substantially off the fairness line. Examination of the Sample 1
pretest plots showed many more items substantially off the
fairness line; the instruction in the class moved female students
nearer the fairness line on many items (except the gender biased
items). This comparison suggests that it is not only the
interactivity of the instruction that matters in reducing the
gender gap but also its overall effectiveness. It seems possible
that the gender gap closes for interactive courses only if they
produce superior learning outcomes, measured by FCI posttest
scores. This could explain the inconsistent relationship
between interactive instruction and lowering the
gender gap
\cite{lorenzo2006reducing,pollock2008comparing,kohl2009introductory,pollock2007reducing}.

Comparing results for Samples 1, 2, and 3 illuminates the
variability of previous research into item
fairness. While not as large as Sample 1, Samples 2 and 3 contain
as many or more students than some of the other studies of item
fairness. Difficulty measures for these samples had large error
bars, particularly for female students. Both samples also involved
confounding factors such as multiple instructors and pedagogies or
a longitudinal application of the FCI which would also increase
variability. The gender biased items were hidden by the noise in
these samples and were probably partially obscured by variation in
other studies. Experiments sub-sampling Sample 1 suggest 1000-1500
as a minimum sample size to clearly resolve gender disparities in
FCI datasets where women are significantly underrepresented.

The inclusion of many biased items calls into question the
practical application of the FCI instrument as well as research
based on the FCI. Examples of the threat to research validity can
be found in two recent studies. In a factor analysis of the FCI
\cite{scott2012exploratory}, gender biased items 21, 22, 23, and
27 factored together while Item 14 failed to be included in any
factor. This raises the question of whether the gender bias of the
questions influenced the factor structure.

Han {\it et al.} \cite{han2015} investigated dividing the FCI into
two shorter tests (half-tests) to lower the time burdens of
testing. Gender fairness was not considered in their analysis.
Randomly, four of the five highly unfair to women questions (14, 21, 22,
and 23) were
included in the second half-test while none of the highly biased
questions were included in the first. The second half-test also included
Item 24 which was identified as unfair after highly unfair items were removed
from the FCI. The first half-test also
contained the two questions that DIF identified as biased toward
women (9 and 15) and two of the additional questions DIF
identified as biased toward men (6 and 12). As such, it is likely
that the second half-test is more gender unfair than the FCI and
the first half-test is more gender neutral.

This study identified a reliable and fair 19-item version
of the FCI. It seems likely, however, that if this instrument were
deployed in diverse educational settings as both a pretest and
posttest that it would produce results with differing levels of
validity for men and women in some situations by posing questions
that are either too easy or too hard for the student population.
As such, instructors using this instrument should be aware of the
possibility of unfairness and either confirm the fairness of the
instrument independently or restrict the kinds of decisions made
from the results of the instrument. For example, using the FCI
pretest as a baseline measurement without instructional
consequences may be appropriate, but using pretest scores to
assign lab groups may not be.

\section{Limitations}

While this research used data from four institutions combined to form three datasets, two of
the datasets were too small to provide adequate
statistical power to determine if some conclusions were general.
The analysis should be conducted with additional large datasets to
determine whether the conclusions are widely replicated.

Additionally, these results suffer from the same methodological
constraints of all large-scale, quantitative studies where binary
gender reporting is used. Coding all students (typically from
institutional records) as male or female simplifies the complexity
of gender identity, ignores the nuances of individual experiences,
and (in the case of DIF) uses male students as the measure of
``normal'' against which female students are compared
~\citep{traxler_enriching_2016}. We chose to replicate these
assumptions for the purpose of engaging with the long tradition of
gender gap studies that follow this model. It is certainly not our
intent to argue that quantitative analysis is the only or the best
method for studying the gendered experiences of students in
learning physics. However, ignoring even this ``first order''
model of gender can lead instructors to base conclusions about
their students on flawed instruments.

\section{Conclusions and Future Work}
\label{sec:conclusions}

The FCI is broadly used to assess physics instruction and
conceptual learning. The above analysis demonstrated that it
contains a number of items that are not fair to women and a few
items unfair to men. The prevalence of the FCI and large
longitudinal datasets that have been collected make it difficult
to suggest that its use should be discontinued; however, the 30-item score
should not be used for any purpose from which a student might
benefit. We suggest the continued use and reporting
of the full FCI score along with the score on the reduced unbiased
instrument. The reduced unbiased instrument score should be used
for instructional decisions and to assign course credit.

The reporting of gender composition is uneven in PER. Researchers
referencing FCI scores at multiple institutions should be aware
that these scores may contain variation that results from  gender
differences that were not reported.

Readers should not attribute our finding of bias in some FCI items
to oversight by the developers of the FCI. By most measures
available to conceptual inventory developers where limited initial
deployment is possible, the FCI performs exceptionally well. The
identification of the unfair items required multiple studies and
very large samples. As such, future developers of conceptual
instruments should plan for a second level of validation which can
only be carried out if their instrument achieves broad deployment.
This validation might identify items with unexpected biases,
reliability, or validity problems. The overall instrument and any
sub-scales should be sufficiently robust that the removal of some
items leaves the validity and reliability of the instrument
intact.

This work will be extended to the FMCE and the CSEM to determine
how much, if any, of the gender gap reported in these instruments
can be attributed to bias. This work should also be extended to
investigate fairness for other underrepresented populations.

\begin{acknowledgments}
This work was supported in part by the National Science Foundation
as part of the evaluation of improved learning for the Physics
Teacher Education Coalition, PHY-0108787.
We appreciate the efforts of the instructors who contributed data to the three samples.
\end{acknowledgments}


%

\end{document}